 \documentclass[12pt]{article}
 \usepackage{graphicx}
 \usepackage{amssymb}
 \usepackage{amsmath}
 \usepackage{amsfonts}
 \usepackage{revsymb}
 \newcommand{\be}{\begin{equation}}
\newcommand{\ee}{\end{equation}}
\newcommand{\bea}{\begin{eqnarray}}
\newcommand{\eea}{\end{eqnarray}}
\newcommand{\bean}{\begin{eqnarray*}}
\newcommand{\eean}{\end{eqnarray*}}

\begin{document}
\begin{titlepage}
\bigskip
%\rightline{}

%%%%%%%%%%%%%%%%%%%%%%%%%%%%%%%%%%%%%%%%%%%%%%%%%%%%%%%%%%%%%%%%%%%%%%%%%%%%%%%%%%%%%%%%%%%%%%%%%%%%%%%%%%%%%%%%%%%%%%%
\rightline {USTC-ICTS-07-18}
\bigskip\bigskip
\bigskip\bigskip

\centerline {\Large \bf {Phase
Transition Dynamics and Its $\alpha^{\prime}$ Corrections}}
\bigskip\bigskip
\bigskip\bigskip

\centerline{\large Yushu Song}
\bigskip\bigskip
\centerline{\em Interdisciplinary Center of Theoretical Studies}

\centerline{\it USTC, Hefei, Anhui 230026, China}

\centerline{\it and}

\centerline{ \it Institute of Theoretical Physics} \centerline{\it
Academia Sinica, Beijing 100080, China} \centerline{\em
yssong@itp.ac.cn}
\bigskip\bigskip
\begin{abstract}
 We study the dynamics of the first order phase transition in the
 holographic hard wall model, namely, Polchinski-Strassler's model and
 come to the conclusion that the phase
 transition is incomplete in large $\mathcal{N} $ limit with the natural
 boundary condition. We also consider the string length corrections to both
  hard wall model and Witten's model, and find that the interesting transition
 configuration is preserved under the $\alpha^{\prime}$ corrections.
\end{abstract}
\end{titlepage}
\baselineskip=16pt

\setcounter{equation}{0}
 \section{Introduction}
The duality proposed by Maldacena \cite{Maldacena:1997re} between
type IIB string theory in AdS$_{5} \times $S$^{5}$ and
$\mathcal{N}=4 $ supersymmetric Yang-Mills theory in four dimensions
has been an arena for amazing theoretical advances in the past
serval years. This duality, often called AdS/CFT duality for
simplicity, has been extended to more realistic situations and shown
to provide a supergravity description of the field theory that
displays confinement and chiral symmetry breaking
\cite{Klebanov:2000hb} \cite{Maldacena:2000yy}. On the other hand,
the fact that black holes have thermodynamic properties is one of
the most striking features of classical and quantum gravity. Thus
studying thermal aspects of AdS/CFT duality will give more insights
into understanding of gauge theory, gravity and Holographic
principle. At finite temperature, Witten \cite{witten-thermal}
generalized this duality to relate the thermodynamics of gauge
theories to that of gravity theories, and to give geometric
descriptions to confining - deconfining phase transitions in certain
gauge theories . The dictionary of the gauge-gravity duality is that
the deconfined phase is described by a black hole, and the confined
phase is described by a nonsingular spacetime without a horizon. On
the gravity side, this transition is the so-called Hawking-Page
phase transition \cite{Hawking:1982dh} and is the first order one.
In \cite{Polchinski:2000uf, Polchinski:2001tt}, Polchinski and
Strassler proposed another interesting phenomenological model by
removing the small radius region of AdS which corresponds to
introducing an IR cutoff in the dual conformal field theory.

Recently, Horowitz and Roberts \cite{Horowitz:2007fe} found an
interesting scenario of the first order phase transition in the
frame of gauge-gravity duality. Starting with supercooled deconfined
phase (with temperature well below the critical temperature), there
are three stages in the phase transition: (i) nucleation of bubbles
of the confined phase, (ii) rapid growth of these bubbles, (iii)
large plasma-ball phase. From then on the evolution depends on the
boundary condition. With a natural description, microcanonical
(fixed energy) description, the plasma-ball is stable and the phase
transition is not complete. The authors of \cite{Horowitz:2007fe}
investigated Witten's model of gauge-gravity duality and found that
in stage (iii) the plasma-ball occupies at least one quarter of the
initial volume.

This note is organized as follows.  In section two we study the
first order phase transition dynamics of Polchinski-Strassler's
model \cite{Polchinski:2000uf, Polchinski:2001tt}, often called hard
wall model, and we get a similar conclusion as that of Horowitz and
Roberts \cite{Horowitz:2007fe}. In section three we pay attention to
the string length corrections of Polchinski-Strassler's model and we
also discuss Witten's model with the $\alpha^{\prime}$ corrections.
Finally in section four we present some discussions and conclusions.

\setcounter{equation}{0}
\section{Gravity analysis of the hard wall model}
Recently, the hard wall model has been studied in many papers
\cite{Herzog:2006ra}-\cite{Cai:2007bq}. Let us first give a brief
review of this model. Following \cite{Maldacena:1997re}, there are
remarkable processes \cite{Polchinski:2000uf, Klebanov:2000hb,
Maldacena:2000yy} in extending the duality to more realistic
conditions in order to give us more insights into QCD and the nature
of confinement. A general realization from these papers is that
confinement is related to supersymmetric field theories whose
gravitational descriptions cap off the geometry in a smooth way in
the infrared at small radius. The gravitational descriptions of
those approaches are cumbersome, but the geometric insight is clear:
IR cutoff of geometry at small radius produces confinement in the
dual gauge theory. Based on this insight, \cite{Polchinski:2001tt}
proposed a far simpler model: the small radius region of AdS is
removed. While such a removal is brutal, subsequent work has shown
that one gets realistic, semi-quantitative descriptions of low
energy QCD \cite{Erlich:2005qh, Da Rold:2005zs}. Some aspects of
thermodynamics of this model have been discussed in papers such as
\cite{Herzog:2006ra, Cai:2007zw}. In this note, dynamics of
confining - deconfining phase transition of this model is concerned.
In the following, we will consider black holes thermodynamics in
AdS, so the five sphere will not play an important role and we will
not write it explicitly in the discussion below. In gauge-gravity
duality, the gravity solution describing the gauge theory at finite
temperature can be obtained by taking the decoupling limit of the
general black 3-brane and keeping the energy density above
extremality finite. Then in the hard wall model the system in the
deconfined phase is described by black 3-brane metric \be
ds^2=\frac{r^2}{L^2}(-f
dt^2+d{x_1}^2+d{x_2}^2+d{x_3}^2)+\frac{L^2}{r^2}f^{-1}dr^2 \ee where
$f=1-\frac{r_0^4}{r^4}$, $L^4=2{g^2_{YM}}N{\alpha^{\prime2}}$, $r_0$
is the mass parameter of the solution, and 't Hooft coupling
constant for the $SU(N)$ gauge theory is
$\lambda=2{g^2_{YM}}N=4{\pi}g_sN$. The black hole horizon is located
at $r=r_0$. The temperature can be determined by Wick rotating to
Euclidean time $\tau=it$ and choosing a periodicity for Euclidean
time $\beta={\pi}L^2/r_0\equiv1/T$ to regularize the singularity at
$r=r_0$. The space only exists for $r>r_{IR}$ due to the hard wall
model with IR cutoff $r_{IR}$. The ground state of confined phase is
described by thermal AdS space \be ds^2=\frac{r^2}{L^2}(-
dt^2+d{x_1}^2+d{x_2}^2+d{x_3}^2)+\frac{L^2}{r^2}dr^2 \label{ads}\ee
with the same IR cutoff $r>r_{IR}$.

According to \cite{Hawking:1982dh}, one can identify the free energy
of the theory with Euclidean gravitational action times temperature,
i.e.\be I=\beta F \ee The 5-dimensional gravitational action
obtained from D=10 type IIB supergravity action by compactifying on
S$^5$ is \be I_5=-\frac{1}{16\pi G_5}\int
dx^5\sqrt{g_5}(\mathcal{R}_5+\frac{12}{L^2})\ee Note that, since
Vol$(S^5)={\pi}^3L^5$, we have \be\frac{1}{16\pi
G_5}=\frac{\pi^3L^5}{16\pi
G_{10}}=\frac{\pi^3L^5}{2\kappa^2}=\frac{\pi^3L^5}{(2\pi)^7{g^2_s}{\alpha^{\prime4}}}
\ee In the following discussions, we will set $L=1$ for simplicity.
Calculation of the contributions from the 2-derivative terms is
divergent at large distance and requires a subtraction. In the black
hole phase, the space exists from $r_{max}=$max$(r_0,r_{IR})$ to
infinity. Therefore, one has the Euclidean action for the black hole
phase \be I_{5b}=-\frac{N^2}{8\pi^2}V_3
\beta_b\int^{R}_{r_{max}}dr\sqrt{g_5}(\mathcal
{R}_5+12)=\frac{N^2}{4\pi^2}V_3\beta_b(R^4-r^4_{max})\ee and
Euclidean action for thermal AdS space phase with IR cutoff
$r>r_{IR}$ \be I_{5a}=-\frac{N^2}{8\pi^2}V_3
\beta_a\int^{R}_{r_{IR}}dr\sqrt{g_5}(\mathcal
{R}_5+12)=\frac{N^2}{4\pi^2}V_3\beta_a(R^4-r^4_{IR})~~.\ee In order
to analyze the dynamics of the first order phase transition, a
natural reference background is the AdS space in the zero limit IR
cutoff. Thus the background contribution is \be
I_{5r}=-\frac{N^2}{8\pi^2}V_3
\beta_r\int^{R}_{0}dr\sqrt{g_5}(\mathcal
{R}_5+12)=\frac{N^2}{4\pi^2}V_3\beta_r R^4~~.\ee We must choose the
appropriate periodicities of $\beta_a$ and $\beta_r$ to ensure that
the solutions can be embedded into the reference background
\be\frac{r}{L}f^{1/2}\beta_b|_{r=R}=\frac{r}{L}\beta_a|_{r=R}
=\frac{r}{L}\beta_r|_{r=R}~~.\ee In the large R limit, we have the
constraint as \be(1-\frac{r^4_0}{2R^4})\beta_b=\beta_a=\beta_r~~.\ee
Subtracting the contribution of the reference background and taking
$R\rightarrow\infty$, one obtains the Euclidean action of the black
hole\be
I^{\prime}_{5b}=\frac{N^2}{4\pi^2}V_3\beta_b(\frac{1}{2}{r^4_0}-r^4_{max})
\label{BHRA}\ee In the light of the results in \cite{Herzog:2006ra,
Cai:2007zw}, that in the classical gravity side, the Hawking-Page
phase transition can occur only when $r_{IR}<r_0$, we take
$r_{max}=r_0$ here. Then the formula (\ref{BHRA}) can be simplified
to \be
I^{\prime}_{5b}=-\frac{N^2}{8\pi^2}V_3\beta_b{r^4_0}=-\frac{N^2\pi^2}{8}V_3{T^3}
\ee Using the
general thermodynamics relations \bea I&=&\beta F\nonumber\\
E&=&\frac{\partial I}{\partial\beta}\\
F&=&E-TS\nonumber\eea we find
\bea F_b&=&\frac{-\pi^2N^2}{8}V_3T^4\nonumber\\
E_b&=&\frac{3\pi^2N^2}{8}V_3T^4\label{bhformula}\\
S_b&=&\frac{\pi^2N^2}{2}V_3T^3\nonumber~~.\eea Taking the same
approach to AdS space (\ref{ads}), we get \bea
I^{\prime}_{5a}&=&-\frac{N^2}{4\pi^2}V_3\beta_b{r^4_{IR}}
\nonumber\\F_a&=&-\frac{N^2}{4\pi^2}V_3{r^4_{IR}}\nonumber\\
E_a&=&-\frac{N^2}{4\pi^2}V_3{r^4_{IR}}\nonumber\\
S_a&=&0~~.\label{adsformula}\eea These energies are measured
relative to the pure thermal AdS space. The AdS space has no horizon
and thus no intrinsic entropy. We can calculate the free energy
difference between the two phases and find: \be
F_b-F_a=\frac{N^2}{8\pi^2}V_3(2r^4_{IR}-r^4_0)~~.\ee So for
$T<T_c=2^{\frac{1}{4}}r_{IR}/\pi$, the thermal gas dominates, and at
$T>T_c=2^{\frac{1}{4}}r_{IR}/\pi$, the black brane dominates. In the
gravitational language, this is the so-called Hawking-Page phase
transition. The above dicussion is just the result of
\cite{Herzog:2006ra}. Now we will consider the dynamics of the phase
transition. Starting with the deconfined phase and quickly lowering
the temperature below $T_c=2^{\frac{1}{4}}r_{IR}/\pi$, how does the
phase transition proceed? Based on the argument of
\cite{Horowitz:2007fe}, bubbles of the confined phase should be
nucleated and grow. On the gravity side, this corresponds to bubbles
of AdS being nucleated on the black 3-brane. At large N limit, the
growth of these bubbles are described by supergravity. We will first
discuss the growth of these bubbles in a microcanonical ensemble
(fixed energy) which is the natural boundary condition for
asymptotically AdS solutions. After the bubbles are nucleated, only
a fraction $\alpha$ of initial volume $V_3$ will be occupied by
black 3-brane and the rest will be AdS space. In this process we
will fix the energy due to the microcanonical ensemble \be E=\alpha
E_b+(1-\alpha)E_a\label{fixenergy} \ee The total energy E is
characterized by the equivalent temperature $T_0$, and the black
brane covering all of $V_3$ will have this total energy.
Substituting the relation (\ref{bhformula}) and (\ref{adsformula})
into (\ref{fixenergy}), we have
\be\frac{3\pi^2N^2}{8}V_3T_0^4=\alpha\frac{3\pi^2N^2}{8}V_3T^4-(1-\alpha)
\frac{N^2}{4\pi^2}V_3{r^4_{IR}}~~.\ee This gives \be
T=[\frac{1}{\alpha}T^4_0+\frac{1-\alpha}{3\alpha}T^4_c]^{\frac{1}{4}}~~.\ee
Thus the entropy of this configuration is just $\alpha$ times the
entropy of the black brane: \bea S&=&\frac{\pi^2}{2}\alpha N^2V_3T^3 \nonumber\\
&=&\frac{\pi^2}{2}
N^2V_3{\alpha}[\frac{1}{\alpha}T^4_0+\frac{1-\alpha}{3\alpha}T^4_c]^{\frac{3}{4}}~~.
\eea The entropy of this configuration is maximized at \be
\alpha=\frac{1}{4}[1+3\frac{T^4_0}{T^4_c}] \ee with $\alpha\in[0,1]$
and $T_0\in[0,T_c]$. One notable feature of this result is that even
as $T_0\rightarrow 0$, the maximal entropy configuration is the
localized black hole taking up a quarter of initial volume. In terms
of gauge theory, there will be a large region of deconfined plasma.
Another interesting feature of this maximal entropy configuration is
that the temperature of the black hole is at critical temperature
independent of initial temperature $T_0$.

\setcounter{equation}{0}
\section{String length corrections to models}
In this section we will focus on contributions from the string
length correction ${\alpha^{\prime3}}\mathcal {R}^4$  to the
supergravity action. In the Einstein frame, using the convention of
including $F^2_5$ in the action and imposing the self duality
constraint on fiveform field strength $F_5=F^*_5$ by hand, the tree
level type IIB string effective action has the following structure:
\be I=-\frac{1}{16\pi G_{10}}\int d^{10}x\sqrt{g}[\mathcal
{R}-\frac{1}{2}(\partial
\phi)^2-\frac{1}{4\cdot5!}F^2_5+\cdots+\gamma
e^{-\frac{3}{2}\phi}W+\cdots] \label{10daction} \ee where \bea
&\gamma=\frac{1}{8}\zeta (3){\alpha}^{\prime3}&\nonumber \\
&W=\mathcal {R}^{hmnk}\mathcal {R}_{pmnq}\mathcal
{R}^{rsp}_{h}\mathcal {R}^{q}_{rsk}+ \frac{1}{2}\mathcal
{R}^{hkmn}\mathcal {R}_{pqmn}\mathcal {R}^{rsp}_{h}\mathcal
{R}^{q}_{rsk}&\nonumber\\&+~~ \mathrm{terms~ depending~ on~ Ricci~
tensor} ~~.&\eea Dots stand for other terms depending on
antisymmetric tensor field strengths and derivatives of dilaton. The
field redefination ambiguity allows one to change the coefficients
of terms involving the Ricci tensor, so there exists a scheme where
$W$ entirely depends on the Weyl tensor \be W=\mathcal
{C}^{hmnk}\mathcal {C}_{pmnq}\mathcal {C}^{rsp}_{h}\mathcal
{C}^{q}_{rsk}+ \frac{1}{2}\mathcal {C}^{hkmn}\mathcal
{C}_{pqmn}\mathcal {C}^{rsp}_{h}\mathcal {C}^{q}_{rsk}
\label{defW}\ee
 The form of $W$ is special in the sense that the
AdS$_{5} \times $S$^{5}$ is still the solution of the action
(\ref{10daction}) with self-dual $F_5$ and constant dilaton. The
${\alpha^{\prime3}}\mathcal {R}^4$ correction to AdS$_5$ black brane
metric has been obtained in \cite{Gubser:1998nz}. Considering
5-dimensional metric \be
ds^2=H^2(K^2d{\tau}^2+P^2dr^2+d{x_1}^2+d{x_2}^2+d{x_3}^2)\ee where
$H$, $K$, and $P$ are functions of $r$ only, then from (\ref{defW})
we have \be
W=\frac{5}{36}\frac{1}{K^4H^8P^4}[(\frac{K^{\prime}}{P})^{\prime}]^4\ee
where primes denote derivatives with respect to $r$. $H$, $K$, and
$P$ are given by \be H=r~~,K=e^{a+4b}~~,P=e^b\ee and \bea a=-2\log
r+\frac{5}{2}\log
(r^4-r^4_0)-\frac{15}{2}\gamma(25\frac{r^4_0}{r^4}+25\frac{r^8_0}{r^8}-79\frac{r^{12}_0}{r^{12}})+\mathcal
{O}({\gamma}^2)\\ b=-\frac{1}{2}\log
(r^4-r^4_0)+\frac{15}{2}\gamma(5\frac{r^4_0}{r^4}+5\frac{r^8_0}{r^8}-19\frac{r^{12}_0}{r^{12}})+\mathcal
{O}({\gamma}^2)\eea The contribution from term $W$ to the action has
been calculated in\cite{Gubser:1998nz} and the result is \be \delta
I=-\frac{1}{16\pi G_5}\int dx^5\sqrt{g_5}\gamma W\ee

\subsection{High order corrections to hard wall model}
Let us first consider the hard wall model with higher order
corrections. In the following discussions, we will consider the
configurations having the confining-deconfining phase transitions,
in which we just take $r_{max}=r_0$. The Euclidean action for the
black hole phase with high order corrections is: \bea
I_{5b}&=&-\frac{N^2}{8\pi^2}V_3
\beta_b\int^{R}_{r_{0}}dr\sqrt{g_5}(\mathcal {R}_5+12+\gamma W)
\nonumber\\&=&-\frac{N^2}{8\pi^2}V_3
\beta_b\int^{R}_{r_{0}}dr[-8r^3+\gamma
(\frac{360r^{16}_0}{r^{13}}+\frac{960r^{12}_0}{r^{9}})+\mathcal
{O}(\gamma^2)]\nonumber\\&=& \frac{N^2}{4\pi^2}V_3\beta_b
[R^4-r^4_0+15\gamma
r^{12}_0(r^4_0(\frac{1}{R^{12}}-\frac{1}{r^{12}_0})+4(\frac{1}{R^{8}}-\frac{1}{r^{8}_0}))]\eea
and for the AdS space with the IR cut-off is:
 \bea I_{5a}&=&-\frac{N^2}{8\pi^2}V_3
\beta_a\int^{R}_{r_{IR}}dr\sqrt{g_5}(\mathcal {R}_5+12+\gamma
W)\nonumber\\&=&\frac{N^2}{4\pi^2}V_3\beta_a(R^4-r^4_{IR})~~.\eea
The natural reference background action takes the form: \be
I_{5r}=\frac{N^2}{4\pi^2}V_3\beta_r R^4~~.\ee Under the
$\alpha^{\prime}$ correction, the periodicity condition becomes
\be[1-\frac{(1+75\gamma)r^4_0}{2R^4}]\beta_b=\beta_a=\beta_r\ee and
the expression of temperature becomes \be
T\equiv\frac{1}{\beta_b}=\frac{(1+15\gamma)r_0}{\pi}~~.\ee Then we
can calculate the thermodynamics quantities of the black hole
phase:\bea
I^{\prime}_{5b}&=&-\frac{N^2\pi^2}{8{\beta^3_b}}V_3(1+15\gamma)\nonumber\\
E_b&=&\frac{3\pi^2N^2}{8\beta^4_b}V_3(1+15\gamma)\\
S_b&=&\frac{\pi^2N^2}{2\beta^3_b}V_3(1+15\gamma)\nonumber\eea and
that of the AdS phase \bea
I^{\prime}_{5a}&=&-\frac{N^2}{4\pi^2}V_3\beta_b{r^4_{IR}}
\nonumber\\
E_a&=&-\frac{N^2}{4\pi^2}V_3{r^4_{IR}}\\ S_a&=&0\nonumber\eea Then
using the same trick of section two, we will easily get the maximal
entropy configuration with
\be\alpha=\frac{1}{4}[1+\frac{3(1+15\gamma)\pi^4T^4_0}{2r^4_{IR}}]
=\frac{1}{4}[1+3\frac{T^4_0}{T^4_c}]\ee and the temperature of this
configuration is just the critical temperature after
$\alpha^{\prime}$ correction, which is a little lower than the
uncorrected one.
\subsection{High order corrections to Witten's model}
In the frame of Witten's model, we will focus on  ${\cal N} = 4$
super Yang-Mills compactified on a circle with antiperiodic
fermions. This breaks the supersymmetry, and gives mass to the
fermions and scalars. The low energy limit is a confining theory,
but not purely 2$+$1 dimensional Yang-Mills theory with regard to
the argument of \cite{Maldacena:2003nj}. Considering ${\cal N} = 4$
super Yang-Mills on ${\bf R}^{3}\times S_\theta^1$, the circle has
antiperiodic fermions, but is spacelike. It does not represent a
Euclidean time direction. The ground state of the confining phase is
described by the AdS soliton \cite{ energyconj} \be
 ds^2=\frac{r^2}{L^2}( -dt^2+dx^2+dy^2+fd\theta^2)+\frac{L^2}{r^2}f^{-1}dr^2
 \ee
 where
 \be\label{deff}
 f=1-\frac{r_s^4}{r^4}
 \ee
The space only exists for $r > r_s$ and regularity at $r=r_s$
requires $r_s=\pi L^2/\beta_{s\theta}$ where $\beta_{s\theta}$ is
the length of $\theta$ direction. At sufficiently high temperatures
$T
> T_c$, the system is in a  deconfined phase  described by the black
3-brane metric \be ds^2=\frac{r^2}{L^2}(-f
dt^2+dx^2+dy^2+d\theta^2)+\frac{L^2}{r^2}f^{-1}dr^2\ee  where
 \be\label{deff}
 f=1-\frac{r_0^4}{r^4}
 \ee
 To avoid the canonical singularity at $r=r_0$, we require $r_0=\pi L^2/\beta\equiv\pi L^2T$
The first order phase transition of this model has been discussed in
\cite{Horowitz:2007fe} by means of the dual classical gravity
analysis. In this subsection, we will discuss the string length
corrections to the dynamics of the phase transitions of Witten's
model.
 Using the same procedure of the above subsection, we have
\bea I_{b}&=&-\frac{N^2}{8\pi^2}V_2
\beta_{\theta}\beta\int^{R}_{r_{0}}dr\sqrt{g_5}(\mathcal
{R}_5+12+\gamma W) \nonumber\\&=&
\frac{N^2}{4\pi^2}V_2\beta_{\theta}\beta [R^4-r^4_0+15\gamma
r^{12}_0(r^4_0(\frac{1}{R^{12}}-\frac{1}{r^{12}_0})+4(\frac{1}{R^{8}}-\frac{1}{r^{8}_0}))]\nonumber\\\eea
for the black hole phase and
 \bea I_{s}&=&-\frac{N^2}{8\pi^2}V_2
\beta_{s\theta}\beta_s\int^{R}_{r_{s}}dr\sqrt{g_5}(\mathcal
{R}_5+12+\gamma W) \nonumber\\&=&
\frac{N^2}{4\pi^2}V_2\beta_{s\theta}\beta_s [R^4-r^4_s+15\gamma
r^{12}_s(r^4_s(\frac{1}{R^{12}}-\frac{1}{r^{12}_s})+4(\frac{1}{R^{8}}-\frac{1}{r^{8}_s}))]\nonumber\\\eea
for the AdS soliton phase. The natural reference background is
taking $r_0, r_s\rightarrow0$ limit, and then we get a thermal AdS
space \bea I_{r}&=&-\frac{N^2}{8\pi^2}V_2
\beta^{\prime}_{\theta}\beta^{\prime}\int^{R}_0dr\sqrt{g_5}(\mathcal
{R}_5+12+\gamma W) \nonumber\\&=&
\frac{N^2}{4\pi^2}V_2\beta_{\theta}\beta
R^4[1-\frac{(1+75\gamma)}{2}\frac{r^4_0}{R^4}+\mathcal
{O}(\frac{r^8_0}{R^8})]\eea In the above derivation, we have used
the periodicity conditions \bea
\beta_s=\beta^{\prime}=\beta[1-\frac{1+75\gamma}{2}\frac{r^4_0}{R^4}+\mathcal
{O}(\frac{r^8_0}{R^8})]
\\\beta_{\theta}=\beta^{\prime}_{\theta}=\beta_{s\theta}[1-\frac{1+75\gamma}{2}\frac{r^4_s}{R^4}+\mathcal
{O}(\frac{r^8_s}{R^8})]\eea After the background subtraction, the
gravitational actions become \bea
I^{\prime}_b=-\frac{N^2\pi^2}{8{\beta^3}}V_2\beta_{\theta}(1+15\gamma)
=-\frac{N^2\pi^2}{8}V_2\beta_{\theta}T^3(1+15\gamma)\label{rbaction}
\\I^{\prime}_s=-\frac{N^2\pi^2}{8{\beta^3_{s\theta}}}V_2\beta_{s}(1+15\gamma)
=-\frac{N^2\pi^2}{8{\beta^3_{s\theta}}T}V_2(1+15\gamma)\label{rsaction}\eea
In order to get(\ref{rsaction}), we have used the $\alpha^{\prime}$
correction to $r_s=\pi L^2/\beta_{s\theta}$ \be (1+15\gamma)r_s=\pi
L^2/\beta_{s\theta}~~. \ee Employing the same procedure as before,
we can calculate the thermodynamics quantities \bea
E_b&=&\frac{3\pi^2N^2}{8\beta^4}V_2\beta_{\theta}(1+15\gamma)\nonumber\\
&=&\frac{3\pi^2N^2}{8}V_2\beta_{\theta}T^4(1+15\gamma)\nonumber\\
S_b&=&\frac{\pi^2N^2}{2\beta^3}V_2\beta_{\theta}(1+15\gamma)\nonumber\\
&=&\frac{\pi^2N^2}{2}V_2\beta_{\theta}T^3(1+15\gamma)\eea
and that of the AdS phase \bea E_a&=&-\frac{\pi^2N^2}{8\beta^3_{s\theta}}V_2(1+15\gamma)\nonumber\\
S_a&=&0\eea Following the derivation of \cite{Horowitz:2007fe}, we
can easily get the maximal entropy configuration with \bea
\alpha&=&\frac{1}{4}[1+3\beta_{\theta}\beta^3_{s\theta}T^4_0]=\frac{1}{4}[1+3\frac{T^4_0}{T^4_c}]\\
T&=&[\beta_{\theta}\beta^3_{s\theta}]^{-\frac{1}{4}}= T_c\eea where
$T_c$ is the phase transition temperature of this process, which can
be obtained by subtracting (\ref{rsaction}) from (\ref{rbaction}).
So the form of $\alpha$ is retained, though the critical temperature
becomes lower under the corrections. The maximal entropy
configuration is the same as that of \cite{Horowitz:2007fe}.

\setcounter{equation}{0}
\section{Discussion and Conclusion}
In this note, we have discussed the dynamics of the first order
confining-deconfining phase transition in the hard wall model.
Following the procedure of \cite{Horowitz:2007fe}, we show a similar
maximal entropy configuration. In the large $\mathcal{N} $ limit,
starting with the supercooled deconfined phase, the process of the
phase transition of the hard wall model with the natural boundary
condition ends up with at least $1/4$ of initial volume of
deconfined phase. This agrees with the conjecture of
\cite{aharony-pballs} that there exists a plasma-ball region that
undergoes the first order phase transition. We also discuss the
string length corrections to the hard wall model and Witten's model.
As we expect, this interesting picture is preserved under the
$\alpha^{\prime}$ corrections.

Besides the models discussed above, there are many other interesting
holographic phenomenological models with confining - deconfining
phase transition phenomenon. Based on the early works
\cite{witten-thermal} \cite{Karch:2002sh}, recently Sakai and
Sugimoto considered the dynamics of flavor D8 branes in the
background geometry of color D4 branes in the probe approximation
\cite{Sakai:2004cn} \cite{Sakai:2005yt} which gives a model of
holographic QCD with matter. The higher derivative corrections to
this model has been studied by Basu \cite{Basu:2007yn}. It is
interesting to study phase transition dynamics in this more
realistic model. There are many models \cite{Behrndt:1998jd}
\cite{AlvarezGaume:2005fv} \cite{Buchel:2007vy} which exhibit
confining - deconfining phase transitions in frame of gauge gravity
duality. The higher derivative corrections of them have been studied
in \cite{Buchel:2006dg} \cite{Dey:2007vt}. That study the phase
transition dynamics of these models to see whether the interesting
picture discussed above is preserved is still an open question.

\vskip 1cm \centerline{\bf Acknowledgements} \vskip .5cm It is a
pleasure to thank Chao-Jun Feng, Tower Wang and Yi Wang for useful
discussions and kind help. We are especially grateful to Miao Li for
a careful reading of the manuscript and valuable suggestions.


\begin{thebibliography}{99}


\bibitem{Maldacena:1997re}
  J.~M.~Maldacena,
  %``The large N limit of superconformal field theories and supergravity,''
  Adv.\ Theor.\ Math.\ Phys.\  {\bf 2} (1998) 231
  [Int.\ J.\ Theor.\ Phys.\  {\bf 38} (1999) 1113]
  [arXiv:hep-th/9711200].
  %%CITATION = IJTPB,38,1113;%%

 %\cite{Klebanov:2000hb}
\bibitem{Klebanov:2000hb}
  I.~R.~Klebanov and M.~J.~Strassler,
  ``Supergravity and a confining gauge theory: Duality cascades and
  chiSB-resolution of naked singularities,''
  JHEP {\bf 0008}, 052 (2000)
  [arXiv:hep-th/0007191].
  %%CITATION = HEP-TH 0007191;%%

  %\cite{Maldacena:2000yy}
\bibitem{Maldacena:2000yy}
  J.~M.~Maldacena and C.~Nunez,
  ``Towards the large N limit of pure N = 1 super Yang Mills,''
  Phys.\ Rev.\ Lett.\  {\bf 86}, 588 (2001)
  [arXiv:hep-th/0008001].
  %%CITATION = HEP-TH 0008001;%%
\bibitem{witten-thermal}
  E.~Witten,
  ``Anti-de Sitter space, thermal phase transition, and confinement in  gauge
  theories,''
  Adv.\ Theor.\ Math.\ Phys.\  {\bf 2}, 505 (1998)
  [arXiv:hep-th/9803131].
  %%CITATION = HEP-TH 9803131;%%


 %\cite{Hawking:1982dh}
\bibitem{Hawking:1982dh}
  S.~W.~Hawking and D.~N.~Page,
  ``Thermodynamics Of Black Holes In Anti-De Sitter Space,''
  Commun.\ Math.\ Phys.\  {\bf 87}, 577 (1983).
  %%CITATION = CMPHA,87,577;%%


%\cite{Polchinski:2000uf}
\bibitem{Polchinski:2000uf}
  J.~Polchinski and M.~J.~Strassler,
  %``The string dual of a confining four-dimensional gauge theory,''
  arXiv:hep-th/0003136.
  %%CITATION = HEP-TH/0003136;%%

%\cite{Polchinski:2001tt}
\bibitem{Polchinski:2001tt}
  J.~Polchinski and M.~J.~Strassler,
  %``Hard scattering and gauge/string duality,''
  Phys.\ Rev.\ Lett.\  {\bf 88}, 031601 (2002)
  [arXiv:hep-th/0109174].
  %%CITATION = PRLTA,88,031601;%%


%\cite{Horowitz:2007fe}
\bibitem{Horowitz:2007fe}
  G.~T.~Horowitz and M.~M.~Roberts,
  %``Dynamics of first order transitions with gravity duals,''
  JHEP {\bf 0702} (2007) 076
  [arXiv:hep-th/0701099].
  %%CITATION = JHEPA,0702,076;%%
%\cite{Erlich:2005qh}
\bibitem{Erlich:2005qh}
  J.~Erlich, E.~Katz, D.~T.~Son and M.~A.~Stephanov,
  %``QCD and a holographic model of hadrons,''
  Phys.\ Rev.\ Lett.\  {\bf 95}, 261602 (2005)
  [arXiv:hep-ph/0501128].
  %%CITATION = PRLTA,95,261602;%%
  %\cite{Da Rold:2005zs}
\bibitem{Da Rold:2005zs}
  L.~Da Rold and A.~Pomarol,
  %``Chiral symmetry breaking from five dimensional spaces,''
  Nucl.\ Phys.\  B {\bf 721}, 79 (2005)
  [arXiv:hep-ph/0501218].
  %%CITATION = NUPHA,B721,79;%%



%\cite{Herzog:2006ra}
\bibitem{Herzog:2006ra}
  C.~P.~Herzog,
  %``A holographic prediction of the deconfinement temperature,''
  Phys.\ Rev.\ Lett.\  {\bf 98}, 091601 (2007)
  [arXiv:hep-th/0608151].
  %%CITATION = PRLTA,98,091601;%%
  %\cite{BoschiFilho:2006pe}
\bibitem{BoschiFilho:2006pe}
  H.~Boschi-Filho, N.~R.~F.~Braga and C.~N.~Ferreira,
  %``Heavy quark potential at finite temperature from gauge / string  duality,''
  Phys.\ Rev.\  D {\bf 74}, 086001 (2006)
  [arXiv:hep-th/0607038].
  %%CITATION = PHRVA,D74,086001;%%
%\cite{Kim:2006ut}
\bibitem{Kim:2006ut}
  Y.~Kim, S.~J.~Sin, K.~H.~Jo and H.~K.~Lee,
  %``Vector Susceptibility and Chiral Phase Transition in AdS/QCD Models,''
  arXiv:hep-ph/0609008.
  %%CITATION = HEP-PH/0609008;%%

%\cite{Kajantie:2006hv}
\bibitem{Kajantie:2006hv}
  K.~Kajantie, T.~Tahkokallio and J.~T.~Yee,
  %``Thermodynamics of AdS/QCD,''
  JHEP {\bf 0701}, 019 (2007)
  [arXiv:hep-ph/0609254].
  %%CITATION = JHEPA,0701,019;%%


%\cite{Kim:2007qk}
\bibitem{Kim:2007qk}
  Y.~Kim, B.~H.~Lee, C.~Park and S.~J.~Sin,
  %``Gluon condensation at finite temperature via AdS/CFT,''
  arXiv:hep-th/0702131.
  %%CITATION = HEP-TH/0702131;%%

%\cite{Kim:2007rt}
\bibitem{Kim:2007rt}
  Y.~Kim, J.~P.~Lee and S.~H.~Lee,
  %``Heavy quarkonium in a holographic QCD model,''
  Phys.\ Rev.\  D {\bf 75}, 114008 (2007)
  [arXiv:hep-ph/0703172].
  %%CITATION = PHRVA,D75,114008;%%
  %\cite{Ballon Bayona:2007vp}
\bibitem{Ballon Bayona:2007vp}
  C.~A.~Ballon Bayona, H.~Boschi-Filho, N.~R.~F.~Braga and L.~A.~Pando Zayas,
  %``On a holographic model for confinement / deconfinement,''
  arXiv:0705.1529 [hep-th].
  %%CITATION = ARXIV:0705.1529;%%
  %\cite{Cai:2007zw}
\bibitem{Cai:2007zw}
  R.~G.~Cai and J.~P.~Shock,
  %``Holographic Confinement/Deconfinement Phase Transitions of AdS/QCD in
  %Curved Spaces,''
  arXiv:0705.3388 [hep-th].
  %%CITATION = ARXIV:0705.3388;%%
  %\cite{Kim:2007em}
\bibitem{Kim:2007em}
  Y.~Kim, B.~H.~Lee, S.~Nam, C.~Park and S.~J.~Sin,
  %``Deconfinement phase transition in holographic QCD with matter,''
  arXiv:0706.2525 [hep-ph].
  %%CITATION = ARXIV:0706.2525;%%

  %\cite{Cai:2007bq}
\bibitem{Cai:2007bq}
  R.~G.~Cai and N.~Ohta,
  %``Deconfinement Transition of AdS/QCD at ${\cal O}(\alpha'^3)$,''
  arXiv:0707.2013 [hep-th].
  %%CITATION = ARXIV:0707.2013;%%
%\cite{Gubser:1998nz}
\bibitem{Gubser:1998nz}
  S.~S.~Gubser, I.~R.~Klebanov and A.~A.~Tseytlin,
  %``Coupling constant dependence in the thermodynamics of N = 4  supersymmetric
  %Yang-Mills theory,''
  Nucl.\ Phys.\  B {\bf 534}, 202 (1998)
  [arXiv:hep-th/9805156].
  %%CITATION = NUPHA,B534,202;%%
%\cite{Maldacena:2003nj}
\bibitem{Maldacena:2003nj}
  J.~M.~Maldacena,
  %``Lectures on AdS/CFT,''
  arXiv:hep-th/0309246.
%%CITATION = HEP-TH/0309246;%%
 \bibitem{energyconj}
  G.~T.~Horowitz and R.~C.~Myers,
  ``The AdS/CFT correspondence and a new positive energy conjecture for
  general relativity,''
  Phys.\ Rev.\ D {\bf 59}, 026005 (1999)
  [arXiv:hep-th/9808079].
  %%CITATION = HEP-TH 9808079;%%
%\cite{aharony-pballs}
\bibitem{aharony-pballs}
  O.~Aharony, S.~Minwalla and T.~Wiseman,
  ``Plasma-balls in large N gauge theories and localized black holes,''
  Class.\ Quant.\ Grav.\  {\bf 23}, 2171 (2006)
  [arXiv:hep-th/0507219].
  %%CITATION = HEP-TH 0507219;%%

  %\cite{Karch:2002sh}
\bibitem{Karch:2002sh}
  A.~Karch and E.~Katz,
  %``Adding flavor to AdS/CFT,''
  JHEP {\bf 0206}, 043 (2002)
  [arXiv:hep-th/0205236].
  %%CITATION = JHEPA,0206,043;%%

  %\cite{Sakai:2004cn}
\bibitem{Sakai:2004cn}
  T.~Sakai and S.~Sugimoto,
  %``Low energy hadron physics in holographic QCD,''
  Prog.\ Theor.\ Phys.\  {\bf 113}, 843 (2005)
  [arXiv:hep-th/0412141].
  %%CITATION = PTPKA,113,843;%%

   %\cite{Sakai:2005yt}
\bibitem{Sakai:2005yt}
  T.~Sakai and S.~Sugimoto,
  %``More on a holographic dual of QCD,''
  Prog.\ Theor.\ Phys.\  {\bf 114}, 1083 (2006)
  [arXiv:hep-th/0507073].
  %%CITATION = PTPKA,114,1083;%%

  %\cite{Basu:2007yn}
\bibitem{Basu:2007yn}
  A.~Basu,
  %``Higher Derivative Corrections in Holographic QCD,''
  arXiv:0707.0081 [hep-th].
  %%CITATION = ARXIV:0707.0081;%%

  %\cite{Behrndt:1998jd}
\bibitem{Behrndt:1998jd}
  K.~Behrndt, M.~Cvetic and W.~A.~Sabra,
  %``Non-extreme black holes of five dimensional N = 2 AdS supergravity,''
  Nucl.\ Phys.\  B {\bf 553}, 317 (1999)
  [arXiv:hep-th/9810227].
  %%CITATION = NUPHA,B553,317;%%

  %\cite{AlvarezGaume:2005fv}
\bibitem{AlvarezGaume:2005fv}
  L.~Alvarez-Gaume, C.~Gomez, H.~Liu and S.~Wadia,
  %``Finite temperature effective action, AdS(5) black holes, and 1/N
  %expansion,''
  Phys.\ Rev.\  D {\bf 71}, 124023 (2005)
  [arXiv:hep-th/0502227].
  %%CITATION = PHRVA,D71,124023;%%

%\cite{Buchel:2007vy}
\bibitem{Buchel:2007vy}
  A.~Buchel, S.~Deakin, P.~Kerner and J.~T.~Liu,
  %``Thermodynamics of the N = 2* strongly coupled plasma,''
  Nucl.\ Phys.\  B {\bf 784}, 72 (2007)
  [arXiv:hep-th/0701142].
  %%CITATION = NUPHA,B784,72;%%

%\cite{Buchel:2006dg}
\bibitem{Buchel:2006dg}
  A.~Buchel,
  %``Higher derivative corrections to near-extremal black holes in type IIB
  %supergravity,''
  Nucl.\ Phys.\  B {\bf 750}, 45 (2006)
  [arXiv:hep-th/0604167].
  %%CITATION = NUPHA,B750,45;%%

%\cite{Dey:2007vt}
\bibitem{Dey:2007vt}
  T.~K.~Dey, S.~Mukherji, S.~Mukhopadhyay and S.~Sarkar,
  %``Phase transitions in higher derivative gravity and gauge theory: R-charged
  %black holes,''
  arXiv:0706.3996 [hep-th].
  %%CITATION = ARXIV:0706.3996;%%











\end{thebibliography}
\end{document}